\documentclass[aps,prd,twocolumn,superscriptaddress,showkeys,showpacs]{revtex4}
\usepackage{pdfsync}
\usepackage{mathptmx}
\usepackage[utf8]{inputenc}
\usepackage[T1]{fontenc}
\usepackage{slashed}
\usepackage{times}
\usepackage{amssymb,amsfonts,amsmath,amsthm}
\usepackage{dsfont,bbm}
\usepackage{dcolumn}
\usepackage{epsf}
\usepackage{graphicx}
\usepackage[caption=false]{subfig}
\usepackage{dsfont}
\usepackage{slashed}
\usepackage[active]{srcltx}
\usepackage[usenames]{color}

\begin{document}

\title{Gluon poles and photon distribution amplitudes in Drell-Yan-like processes}

\author{I.~V.~Anikin}
\email{anikin@theor.jinr.ru}
\affiliation{Bogoliubov Laboratory of Theoretical Physics, JINR,
             141980 Dubna, Russia}
\author{L.~Szymanowski}
\affiliation{National Centre for Nuclear Research (NCBJ),
            00-681 Warsaw, Poland}

\begin{abstract}
We calculate the hadron tensor related to the photon-induced Drell-Yan process,
with incoming nucleon being transversally polarised.
We predict the new single spin asymmetry which probes gluon poles
and which is expressed in terms of
photon (both chiral-odd and chiral-even) distribution amplitudes and chiral-odd nucleon function
stemming from the nucleon transverse polarization.
\end{abstract}
\date{\today}
\maketitle

\section{Introduction}

In experimental studies, the single spin asymmetry (SSA) is a useful tool for such kind of investigations.
In particular, the single transverse spin asymmetry gives a lot of information on the three-dimensional nucleon structure owing to
the nontrivial connection between the nucleon transverse spin and the transverse momentum dependence of parton distributions
(see, for example, \cite{Angeles-Martinez:2015sea, Boer:2011fh, Boer:2003cm, Kang:2011hk, Boer:2011fx, Arnold:2008kf}).
Among the semi-inclusive reactions, the Drell-Yan-like (DY-like) processes play an unique role due to a possibility
to uncover the gluon pole contributions and to study finest structure of hadron transverse polarization effects
\cite{Anikin:2010wz, Anikin:2015xka, Anikin:2015esa, Anikin:2017azc, Anikin:2017vvn, Anikin:2017pch}.

Moreover, there are several experimental projects which pursue the measurements with Drell-Yan-like processes,
RHIC \cite{Bai:2013plv}, COMPASS \cite{Baum:1996yv, COMPASS} and future NICA \cite{Kouznetsov:2017bip, Savin:2016arw}.

As well-known, the DY-like processes with an essential transverse polarization of nucleons
give a possibility to explore the gluon poles
in the twist-$3$ quark-gluon parton distribution functions (PDFs), denoted as $B^V$-PDFs in, for instance, \cite{Anikin:2010wz, Anikin:2015xka, Anikin:2015esa}.
In Ref.~\cite{Anikin:2017azc}, based on the duality properties, the twist-$3$ $B^V$-PDF has been
modelled by the corresponding convolution of two hadron distribution amplitudes.
This model has allowed to study the gluon pole presence in more detail and it has resulted in the prediction of
new SSAs related to the pion-induced DY process which are reachable for measurements by COMPASS \cite{COMPASS}.

Here, we propose a new study of the DY-like processes where the gluon poles lead in significant ways to to new SSA observables.
In particular, we would like to focus our attention on the photon-induced DY process where
the twist-$3$ photon distribution amplitudes (photon DAs)
can be probed. Let us emphasise that the understanding of hadron-like behavior of photons, which is encoded in
the photon DA, still attracts the attention of community \cite{Ioffe:1983ju, Balitsky:1989ry, Ball:2002ps, Braun:2002en, Berger:1979du}.
The attempt to study the leading twist-$2$ photon DA was implemented some years ago in \cite{Pire:2009ap}.

In this paper we calculate the Drell-Yan hadron tensors related to the photon-nucleon processes with the essential
spin transversity and ``primordial'' transverse momenta where the gluon poles have been shown up.
We construct a new single spin asymmetry the existence of which entirely depends on the presence of gluon poles.
We dwell on the case where one of photon DAs has been projected onto the twist-$2$ chiral-odd structure, $\langle \bar\psi \sigma^{+\perp} \psi \rangle$.
We stress that the photon chiral-odd DA
singles out the chiral-odd tensor combination in the nucleon matrix element with essential transverse polarization.
Notice that, in this case, the access to the specific angular dependence of SSA (see Eqn.~(\ref{SSA})) expressed in terms of
the Collins-Soper angles has been opened only owing to the gluon pole presence.
In other words, the experimental verification of SSA’s angular dependence predicted in the present paper
can give evidences for the gluon pole probing and the hadron-like behaviour of photons.

\section{Kinematics}

Let us now go over to the discussion of kinematics.
We consider the photon-induced Drell-Yan process
with nucleon being transversely polarized at the regime where Bjorken fraction $x_B$
is close to $1$, {\it i.e.} $x_B\to 1$ :
\begin{eqnarray}
\label{Process}
\gamma(P_1) + N^{(\uparrow\downarrow)}(P_2) &\to& \gamma^*(q)+ \bar q(K) + X(P_X)
\\
&\to& \bar q(K) + \ell(l_1)+\bar\ell(l_2) + X(P_X).
\nonumber
\end{eqnarray}
The virtual photon producing the lepton pair ($l_1+l_2=q$) has a large mass squared ($q^2\approx Q^2$)
while its transverse momenta $q_\perp$ are small and integrated out.

The light-cone (Sudakov) decompositions supplemented with usual approximations take the forms \cite{Anikin:2017azc}
\begin{eqnarray}
\label{Sudakov-decom-1}
&&P_1=\frac{Q}{x_B \sqrt{2}}\, n^\star
- \frac{x_B P^2_{1\,\perp}}{Q \sqrt{2}} n
+ P_{1\,\perp}
\approx \frac{Q}{x_B \sqrt{2}}\, n^\star  + P_{1\,\perp} \, ,
\nonumber\\
&&P_2\approx \frac{Q}{y_B \sqrt{2}}\, n  + P_{2\,\perp} \, ,
\quad S\approx \frac{\lambda}{M_2}\,P_2 + S_\perp\,
\end{eqnarray}
for the real photon and nucleon momenta and nucleon spin vector;
\begin{eqnarray}
\label{Sudakov-decom-2}
q=\frac{Q}{\sqrt{2}}\, n^\star +\frac{Q}{\sqrt{2}}\,n + q_{\perp},\, \quad q_\perp^2\ll Q^2\, ,
\end{eqnarray}
for the virtual photon momentum.
The momenta $P_1$ and $P_2$ have the plus and minus dominant light-cone
components, respectively.

It is convenient to define the Collins-Soper frame, {\it i.e.} the center-mass-system of lepton pair, as \cite{Barone:2001sp}
\begin{eqnarray}
\label{CS-frame}
&&\widehat{t}=\frac{q}{Q},\quad \widehat{x}=\frac{q_{\perp}}{Q_\perp},\quad
\widehat{z} = \frac{x_B}{Q} \widetilde{P}_1 - \frac{y_B}{Q} \widetilde{P}_2,
\end{eqnarray}
where $\widetilde{P}_1= {P}_1 - q/(2x_B)$ and $\widetilde{P}_2 = {P}_2 - q/(2y_B)$
\footnote{We are not precise about writing the covariant and contravariant vectors in any kinds of
definitions and summations over the four-dimensional vectors, except the cases where this notation may lead to misunderstanding.}.
Moreover, we have
\begin{eqnarray}
\label{lcb}
\sqrt{2} n^\star = \widehat{t} + \widehat{z} - \frac{Q_\perp}{Q} \widehat{x},
\quad
\sqrt{2} n = \widehat{t} - \widehat{z} - \frac{Q_\perp}{Q} \widehat{x}.
\end{eqnarray}

\section{Hadron tensor}

The processes (\ref{Process}) are studied in the kinematics with a large $Q^2$, so we apply the
factorization theorem to get the corresponding hadron tensor factorized in the form
where the hard part is convoluting with the soft part.
All important steps of our factorization approach can be found, for instance, in
\cite{Anikin:2010wz, Anikin:2015xka, Anikin:2015esa} (see, also the references therein).

\begin{figure}[ht]
\vspace{0.3cm}
\centerline{\includegraphics[width=0.4\textwidth]{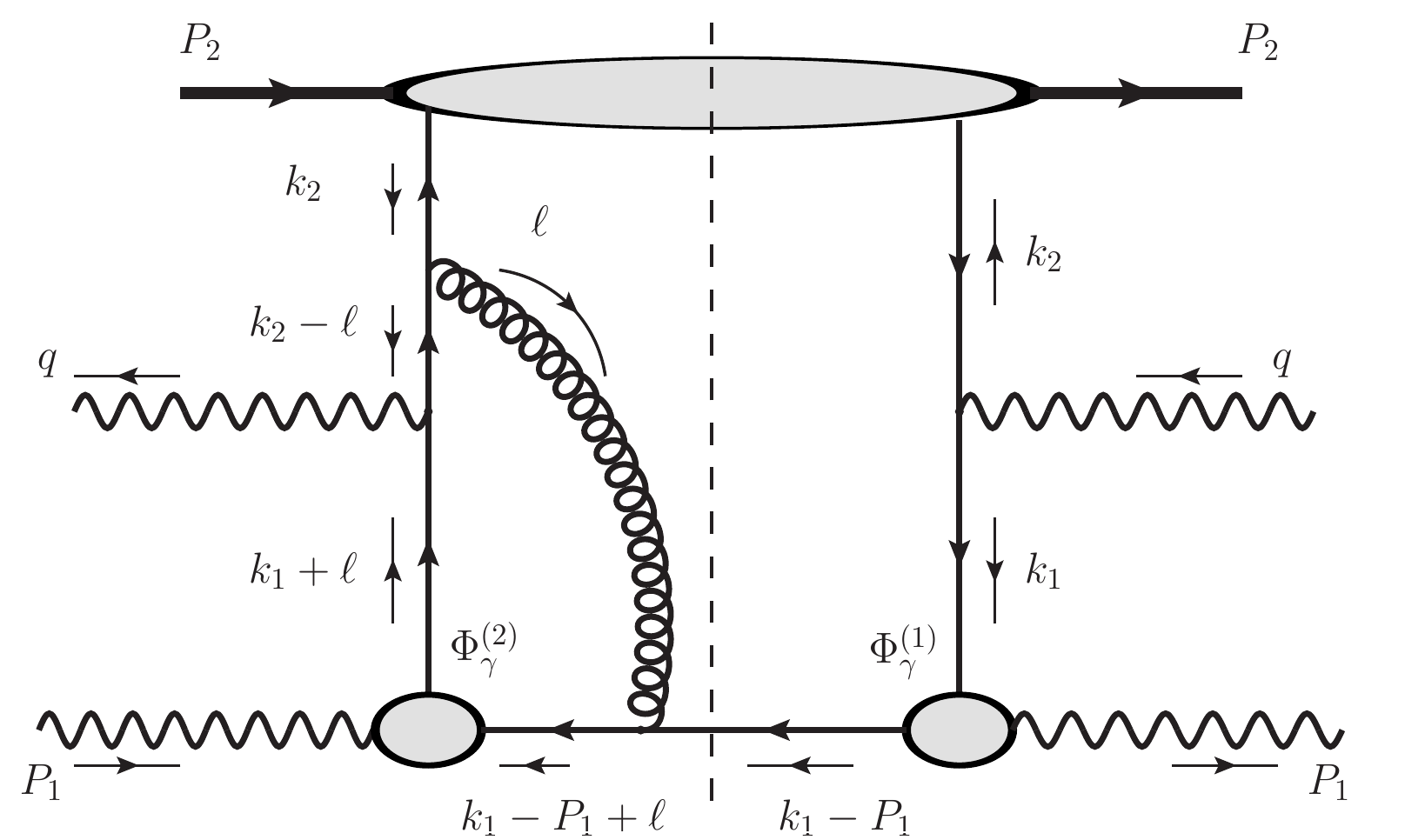}}
\vspace{0.3cm}
\centerline{\includegraphics[width=0.4\textwidth]{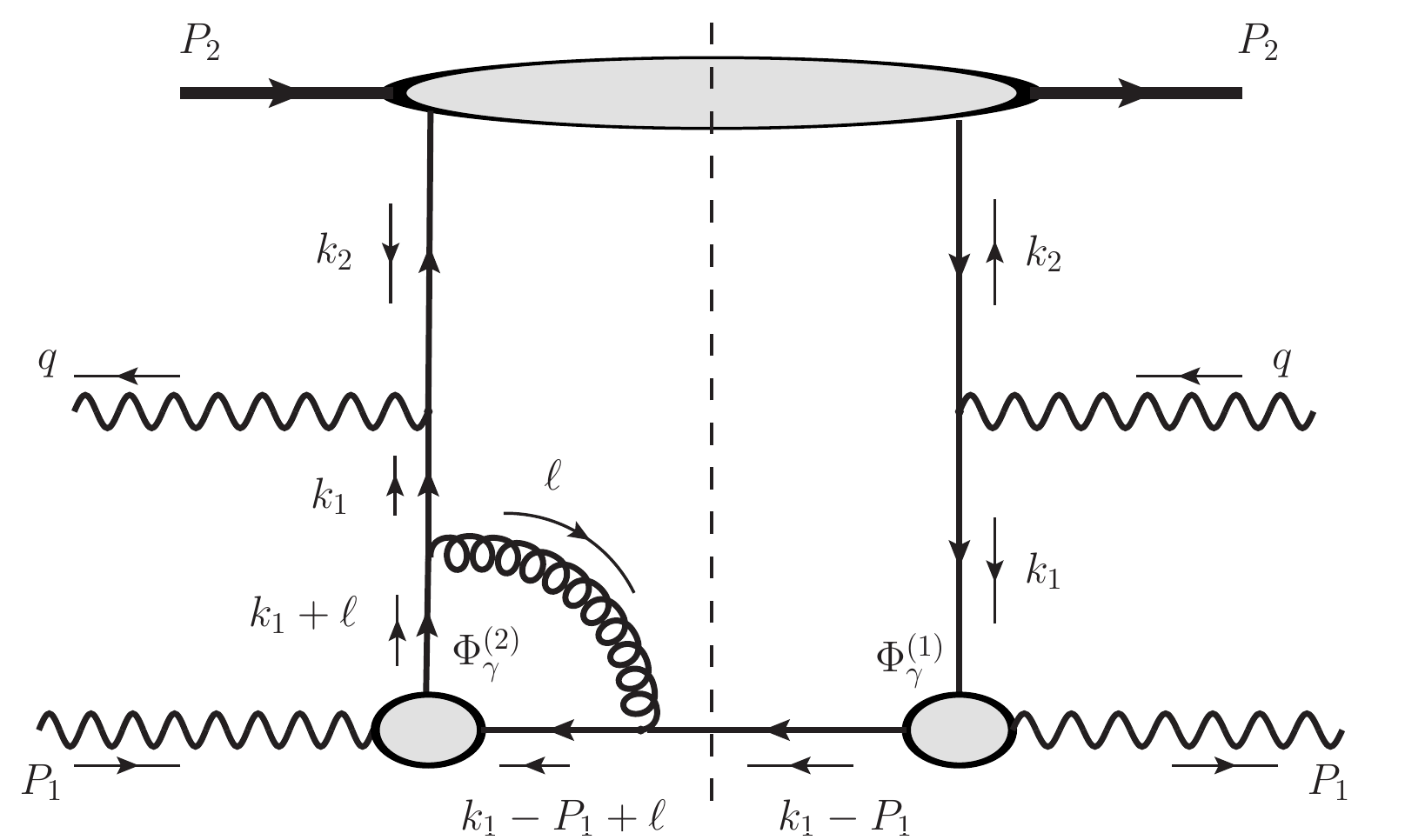}}
  \caption{The ``standard'' diagram (the upper panel) and
  the ``non-standard'' diagram (the lower panel) do not contribute to
  the considered hadron tensor of photon-induced DY process.}
\label{Fig-DY-1-2-b}
\end{figure}

\begin{figure}[ht]
\vspace{0.2cm}
\centerline{
\includegraphics[width=0.4\textwidth]{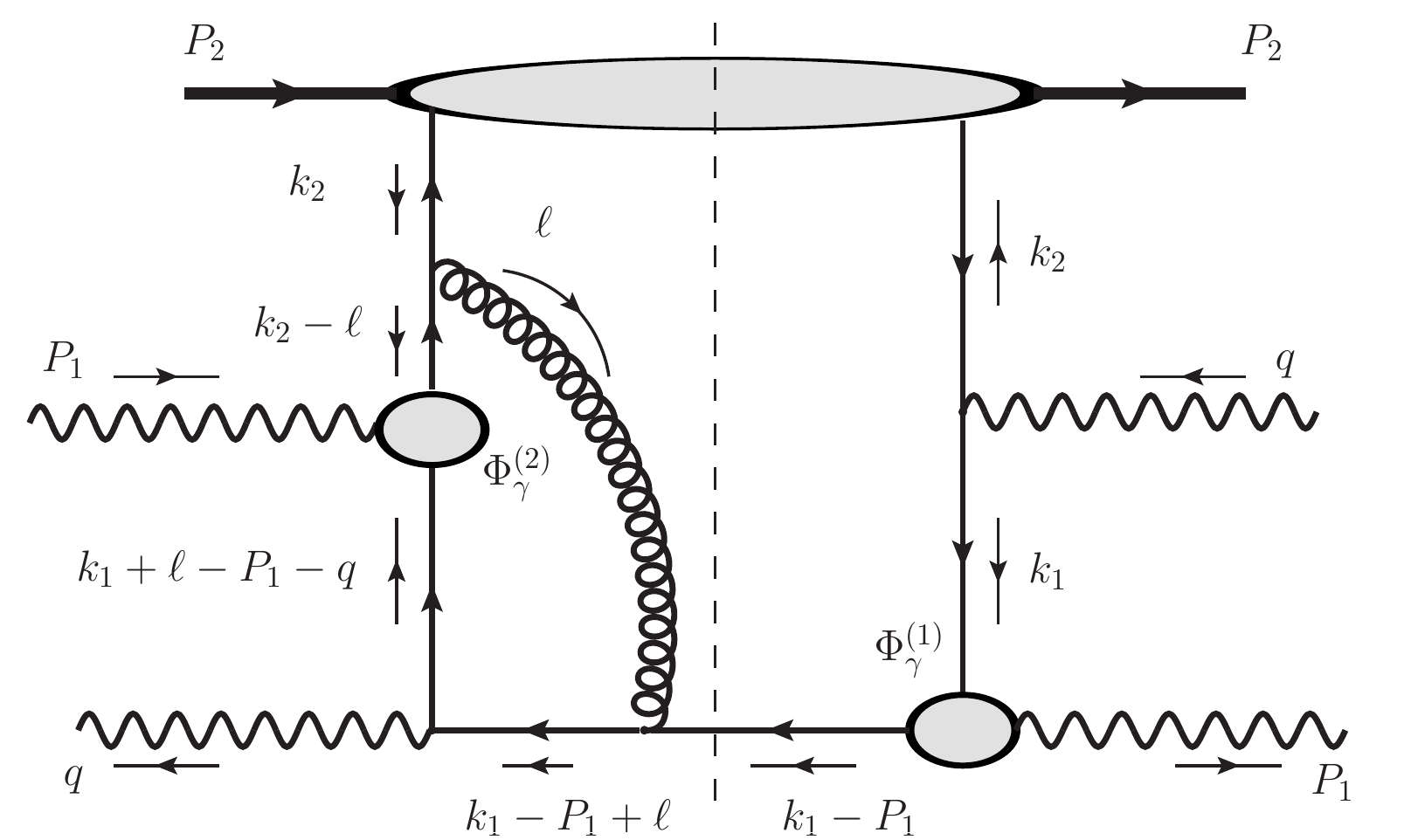}}
  \caption{The Feynman diagrams which contribute to the photon-induced Drell-Yan hadron tensor.}
\label{Fig-DY-1-2-a}
\end{figure}

It is natural to study the role of observables at twist-$3$ level by exploring of different associated asymmetries.
Any single spin asymmetries (SSAs) can be presented in the following symbolical form
\begin{eqnarray}
\label{SSA-symbol}
{\cal A} \sim d\sigma^{(\uparrow)} - d\sigma^{(\downarrow)}
\sim {\cal L}_{\mu\nu}\, H_{\mu\nu}\, ,
\end{eqnarray}
where ${\cal L}_{\mu\nu}$ is the corresponding leptonic tensor and
$H_{\mu\nu}$ stands for the hadronic tensor.
For our purposes, it is enough to consider the case of unpolarized leptons
which
leads to the real lepton tensor and, therefore, the hadron tensor $H_{\mu\nu}$ has to be real one too.
As for any DY-like processes, the hadron tensor includes at least two non-perturbative blobs
which are associated with two different dominant (plus and minus) light-cone directions.
For the process (\ref{Process}), the hadron tensor involves
the upper non-perturbative blob (see Figs.~\ref{Fig-DY-1-2-b} and \ref{Fig-DY-1-2-a})
which corresponds to the matrix elements with the spin transversity:
\begin{eqnarray}
\label{UpperDY}
&&\langle P_2, S_\perp | \bar\psi\, \sigma^{-\perp} \,
\psi |S_\perp, P_2 \rangle\stackrel{{\cal F}}{\sim} \varepsilon^{- \perp S^{\perp} P_{2}}
\, \bar h.
\end{eqnarray}
Notice that the upper blob corresponds to the dominant minus light-cone direction and
the dominant antiquark momentum $k_2$ is defined along the minus direction too.
In more general case, the function $\bar h(y)$ can be expressed through the
corresponding moments of the transverse momentum dependent functions.

In contrast to the usual DY-process (with two initial nucleons) with the twist-$3$ $B^V$-PDF \cite{Anikin:2010wz},
in the photon-induced DY case the lower non-perturbative blob splits
into two photon distribution amplitudes: the one of twist-$2$, $\langle \bar\psi\,\sigma^{+\perp} \psi \rangle$,
and the second one of twist-$3$, $\langle \bar\psi\,\gamma^\perp \psi \rangle$ (in the similar way as for the pion-induced DY-process
\cite{Anikin:2017azc}).

Generally speaking, the hadron tensor involves the contributions from two different
kinds of amplitude interference,
see Figs.~\ref{Fig-DY-1-2-b} and \ref{Fig-DY-1-2-a}, together with the mirror contributions.
Moreover, the hadron tensor term which arises from such a interference
contains the contributions from both the ``standard'' and ``non-standard'' diagrams as shown in Fig.~\ref{Fig-DY-1-2-b}
\footnote{The definitions of ``standard'' and ``non-standard'' diagrams can be found in \cite{Anikin:2010wz, Anikin:2017azc}.}.
As was explained in \cite{Anikin:2016bor}, in this case the gluon propagator involves the transverse gluons only,
{\it i.e.} with the the gluon propagator numerator $d_{\alpha\beta}(\ell)=g^\perp_{\alpha\beta}$.
Thus, it is easy to see, by simple $\gamma$-algera, that for the lower blobs the convolution combination
of $\langle \bar\psi\,\sigma^{+\perp} \psi \rangle$ and $\langle \bar\psi\,\gamma^\perp \psi \rangle$ as presented
in Fig.~\ref{Fig-DY-1-2-b} does not contribute to the hadron tensor at all.
%

Hence, we are left with the only possible interference which generates the hadron tensor shown in Fig.~\ref{Fig-DY-1-2-a}
which refers to the ``standard'' diagram. Note that the “non-standard” diagram vanishes in this case.

We are now in position to discuss the derivation of hadron tensor generated by the diagram in Fig.~\ref{Fig-DY-1-2-a}.
As mentioned, we adhere the terminology and the {\it collinear} factorization procedure as described
in \cite{Anikin:2010wz, Anikin:2015xka, Anikin:2015esa}.
The dominant quark and gluon momenta $k_1$ and $\ell$ lie
along the plus direction while, as above-mentioned, the dominant antiquark momentum $k_2$ -- along the minus direction
(see Figs.~\ref{Fig-DY-1-2-b} and \ref{Fig-DY-1-2-a}). That is the behaviours of parton momenta on the light-cone
satisfy the conditions:
\begin{eqnarray}
k_1\,\, \text{and}\,\, \ell\sim\left(P_1^+,\, \frac{\Lambda_1^2}{P^{+}_1},\, \Lambda_1 \right), \quad
k_2\sim \left(\frac{\Lambda_2^2}{P^{-}_2}, P^-_2, \Lambda_2 \right),
\end{eqnarray}
where $\Lambda_i$ characterize the hadron typical scales.

Before factorization, this diagram leads to the following contribution
(all prefactors are included in the integration measures)
\begin{eqnarray}
\label{HadTen-St}
&&{\cal W}_{\mu\nu}=\int (d^4k_1)\,(d^4k_2) \, \delta^{(4)}(k_1+k_2-q)
\nonumber\\
&&\times\int (d^4\ell)\, {\cal D}_{\alpha\beta}(\ell)
\text{tr}\big[ \gamma_\nu \sigma^{+\perp} \gamma_\alpha \sigma^{-\perp} \gamma_\mu S(k_1+\ell-P_1) \gamma_\beta \gamma^\perp
\big]
\nonumber\\
&&\times\bar\Phi^{[\sigma^{-\perp}]}(k_2)\,
\Phi_{(2)\gamma}^{[\sigma^{+\perp}]}(k_2; \ell)\,
\Phi_{(1)\gamma}^{[\gamma^\perp]}(k_1)\, \delta\big( (P_1-k_1)^2 \big) +
\nonumber\\
&&(\text{Fierz projection replacement}: [\sigma^{+\perp}]\leftrightarrow [\gamma^\perp])
\,,
\end{eqnarray}
where
\begin{eqnarray}
\label{GluonProp}
{\cal D}_{\alpha\beta}(\ell)= \Delta(\ell)d_{\alpha\beta}(\ell), \quad \Delta(\ell)=\frac{1}{\ell^2+i0},
\end{eqnarray}
and
\begin{eqnarray}
\label{barPhi-func}
&&\bar\Phi^{[\sigma^{-\perp}]}(k_2)= \int\hspace{-0.4cm}\sum\nolimits_{X}\,\int (d^4\eta_2)\, e^{-ik_2\eta_2}\,
\\
&&\times\langle P_2, S_\perp |\text{tr}\big[ \psi(0)|P_{X}\rangle \, \langle P_{X} | \bar\psi(\eta_2) \sigma^{-\perp} \big] | S_\perp , P_2\rangle\,
\nonumber
\end{eqnarray}
and
\begin{eqnarray}
\label{Phi2-func}
&&\Phi_{(2)\gamma}^{[\sigma^{+\perp}]}(k_2; \ell)=i\,P_1^+ e_{(\lambda)}^{\perp} \varphi^{(2)}_{\gamma}(k_2,\ell)=
\\
&& \int (d^4\eta_1)\, e^{i(P_1+k_2-\ell)\eta_1}\, \langle 0 | \bar\psi(\eta_1) \,\sigma^{+\perp} \psi(0) | P_1\rangle\,,
\nonumber\\
\label{Phi1-func}
&&\Phi_{(1)\gamma}^{[\gamma^\perp]}(k_1)=e_{(\lambda)}^{\perp} \varphi^{(1)}_{\gamma}(k_1)=
\\
&&\int (d^4\xi)\, e^{-ik_1\xi}\,
\langle P_1 | \bar\psi(\xi) \,\gamma^\perp \psi(0) | 0\rangle\,.
\nonumber
\end{eqnarray}
In Eqns.~(\ref{Phi2-func}) and (\ref{Phi1-func})
the factors which include the quark condensate $\langle\bar\psi \psi\rangle$, the magnetic susceptibility $\chi$, and
the non-perturbative constant $f_{3\gamma}$ are absorbed in the definitions of $\varphi^{(i)}_{\gamma}$ (see \cite{Ball:2002ps} for details).
In  Eqn.~(\ref{HadTen-St}) we write explicitly the $\delta$-function which shows that
the quark operator with the momentum $P_1-k_1$ corresponds to the on-shell fermion.
Moreover, since we here deal with the only ``standard'' diagram, within the
contour gauge $[z^-;-\infty^-]_{A^+}=1$ (see, \cite{Anikin:2010wz, Anikin:2015xka, Anikin:2015esa})
which leads to the local axial gauge $A^+=0$,
the gluon propagator in Eqn.~(\ref{HadTen-St}) contains contributions of two contractions:
$\langle A^\perp A^\perp\rangle$ and $\langle A^\perp A^-\rangle$.
In other words, the numerator $d^{\alpha\beta}(\ell)$ in Eqn.~(\ref{GluonProp}) receives the
contributions from $d^{\perp\,\perp}(\ell)$ and $d^{\perp\, -}(\ell)$ combinations.

The next step is to perform the factorization procedure for the hadron tensor.
We are not going to discuss details of factorization stages
(the comprehensive description of factorization can be found
in many papers, see, for example, \cite{An-ImF, Belitsky:2005qn, Diehl:2003ny, Braun:2011dg}).

By referring to \cite{Anikin:2017azc}, in the limit $x_B\to 1$, we finally derive the following expression
for the gauge invariant hadron tensor (here, $x_{21}=x_2-x_1$):
\begin{eqnarray}
\label{HadTen-GI}
&&\hspace{-0.3cm}\overline{\cal W}_{\mu\nu}= \int d^2\vec{\bf q}_\perp {\cal W}_{\mu\nu}=
\\
&&\hspace{-0.3cm}=
i\,\frac{s^2}{4}\int (dx_1)\,(dy) \, \delta(x_1P^+_1-q^+)\delta(yP^-_2-q^-) \Phi^{(1)}_\gamma(x_1)
\nonumber\\
&&\hspace{-0.3cm}\times \int(dx_2)
\int(dk_2^+ d^2\vec{\bf k}_{2}^{\perp}) \bar h(y| k_2^+,\vec{\bf k}_{2}^{\perp})
\Phi^{(2)}_\gamma(y, x_{21}| k_2^+,\vec{\bf k}_{2}^{\perp}) \,
\nonumber\\
&&\hspace{-0.3cm}\times
\frac{1}{2x_{21}} \left[ g_{\mu\nu}^\perp\, (e^{(\lambda)}_\perp\cdot e^{*\,(\lambda)}_\perp)
-e^{(\lambda)}_{\perp\,\nu}\,e^{*\,(\lambda)}_{\perp\,\mu} - e^{(\lambda)}_{\perp\,\mu}\,e^{*\,(\lambda)}_{\perp\,\nu}\right]
\varepsilon^{P_{1}^{\perp} S_\perp n^\star n}
,
\nonumber
\end{eqnarray}
where
\begin{eqnarray}
\label{Phi1}
&&\Phi^{(2)}_\gamma(y, x_{21}| k_2^+,\vec{\bf k}_{2}^{\perp})=
\int(d\ell^- d^2\vec{\ell}_{\perp})
\frac{\vec{\ell}_\perp\cdot\vec{\bf P}_{1}^{\perp}}{(\vec{\bf P}_{1}^{\perp})^2}
\\
&&\times \Phi^{(2)}_\gamma(y, x_{21}| k_2^+,\vec{\bf k}_{2}^{\perp}; \ell^-, \vec{\ell}_{\perp})
\Delta(\ell^-, \vec{\ell}_\perp)\Delta(\ell^-, \vec{\ell}_\perp - \vec{\bf P}_{1}^{\perp})
\nonumber
\end{eqnarray}
and
$\Phi^{(i)}_{\gamma}(..)=\varphi^{(i)}_{\gamma}(..)/(2\bar x_i P_1^+)$.

In Eqns.~(\ref{HadTen-GI}) and (\ref{Phi1}) we distinguish the arguments of functions
$\Phi^{(2)}_\gamma$ and $\bar h$ in the dominant light-cone variables ($y, x_{21}$) and
(sub)sub-dominant variables ($k_2^+,\vec{\bf k}_{2}^{\perp}; \ell^-, \vec{\ell}_{\perp}$).

\begin{figure}[ht]
\vspace{0.2cm}
\centerline{
\includegraphics[width=0.2\textwidth]{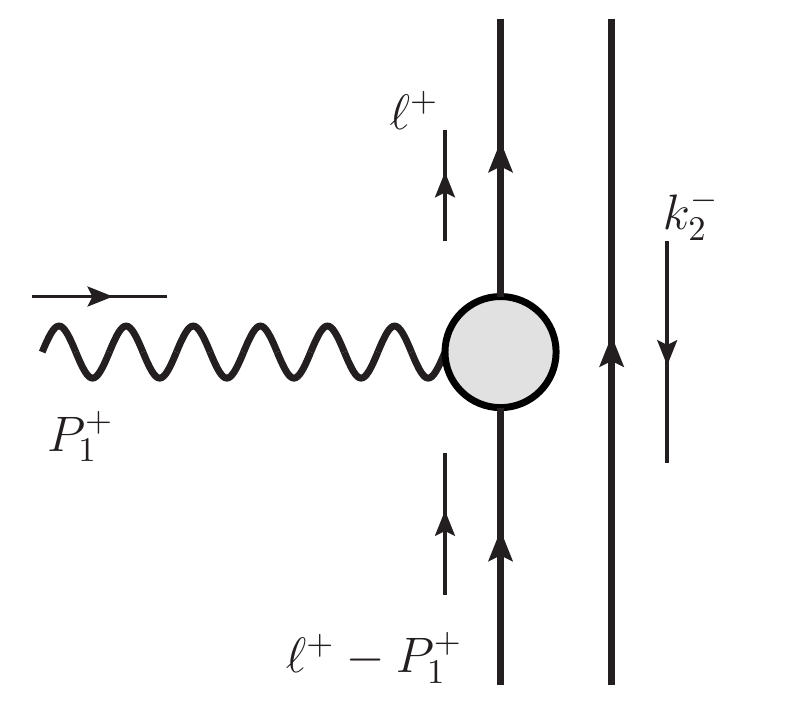}}
  \caption{The longitudinal separable ansatz illustration.}
\label{Fig-3}
\end{figure}

Let us now discuss the possible simplifications which can be applied for both
the longitudinal and transverse momentum dependences of function $\Phi^{(2)}_\gamma$.
First, since this function is parametrizing
the hadron matrix element with the photon carrying the momentum $P_1$ with the plus light-cone
dominant component, it is clear that the momentum $k_2$ with the minus light-cone
dominant component does not affect the dynamics of quark-(real)photon interaction.
So, the simplest case of longitudinal momentum separation has been illustrated in Fig.~\ref{Fig-3}.
The transverse momentum dependence separation, however, is more conditional.
Within our frame, it is natural to assume that $|\vec{\ell}_\perp^{\,2}|\sim |\vec{\ell}_\perp\cdot\vec{\bf k}_2^\perp|
\sim |\vec{\bf k}_{2\,\perp}^{\,2}|\sim O(1)$. Then, we can impose the separable ansatz in the form
\begin{eqnarray}
\label{ansatz}
&&\Phi^{(2)}_\gamma(y, x_{21}| k_2^+,\vec{\bf k}_{2}^{\perp}; \ell^-, \vec{\ell}_{\perp})
\nonumber\\
&&
\approx
\widetilde\Phi^{(2)}_\gamma(x_{21}| \vec{\ell}_{\perp})
\varphi_S(y| \vec{\bf k}_{2}^{\perp})+ O\left( \frac{\ell^-}{P_1^+}, \frac{k_2^+}{P_2^-}\right),
\end{eqnarray}
where $\varphi_S$ has a role of ``spectator'' function.
Justification of the separability in Eqn.~(\ref{ansatz}) can be demonstrated
by using of the effective Lagrangian method in order to calculate the
non-perturbative photon-vacuum matrix element of twist-$2$ quark operators
which corresponds to $\Phi^{(2)}_\gamma$, see Appendix~\ref{App1}.

Finally, we focus on discussion of the pole contribution $1/x_{21}$ in Eqn.~(\ref{HadTen-GI}).
Similarly as in Ref.~\cite{Anikin:2017azc},
the gluon pole, which has to be treated within the contour gauge framework, can be described
as \cite{Anikin:2010wz, Anikin:2015xka, Anikin:2015esa}
\begin{eqnarray}
\label{GP}
\frac{1}{x_2-x_1} \stackrel{\text{c. g.}}{\Longrightarrow}
\frac{1}{x_2-x_1 - i\epsilon}.
\end{eqnarray}
The complex prescription originates from the corresponding integral representation of
theta function (see, \cite{Anikin:2015xka} for details).
This complex prescription in (\ref{GP}) generates the imaginary part in order
to compensate the complex $i$ in parametrization of the photon-vacuum matrix element (\ref{Phi2-func})
and which leads to real part on the {\it l.h.s.} of (\ref{SSA-symbol}).

\section{Single Spin Asymmetry}

We now construct the single spin asymmetry observable.
Within the Collins-Soper frame \cite{Arnold:2008kf, Barone:2001sp},
performing the invariant integration in (\ref{HadTen-GI}) and
contracting the leptonic and hadron tensors, we finally obtain
(we remind that ${\cal L}_{\mu\nu}$ corresponds the unpolarized lepton tensor)
\begin{eqnarray}
\label{SSA}
&&\hspace{-0.6cm}\frac{4\,{\cal L}_{\mu\nu}\, \Im\text{m}\overline{\cal W}_{\mu\nu}}{Q^2} =
\Phi^{(1)}_{\gamma}(x_B)\,\widetilde\Phi^{(2)}_{\gamma}(\bar x_B)
\bar h(y_B)\hspace{-0.2cm} \stackrel{\hspace{0.2cm}\vec{\bf k}_{2}^{\,\perp}}{\circledast}
\hspace{-0.1cm}\varphi_{S}(y_B)
\\
&&\hspace{-0.6cm}\times
\frac{2}{x_B\,y_B}(e^{(\lambda)}_\perp\cdot e^{*\,(\lambda)}_\perp) P_1^\perp \wedge S_\perp
\,\sin^2\theta\cos 2\varphi,
\nonumber
\end{eqnarray}
where we use  Eqn.~(\ref{ansatz}) for $\Phi^{(2)}_\gamma$ and
$\hspace{-0.2cm} \stackrel{\hspace{0.2cm}\vec{\bf k}_{2}^{\,\perp}}{\circledast}
\hspace{-0.1cm}$ stands for the corresponding integration over $\vec{\bf k}_{2}^{\,\perp}$ (cf. Eqn.~(\ref{HadTen-GI})).

Therefore, we predict a new asymmetry which reads (cf. \cite{COMPASS})
\begin{eqnarray}
\label{NewSSA}
&&{\cal A}_T=\frac{S_{\perp}}{Q} \, \frac{  D_{[\theta, \varphi]} \, \sin(\phi_S+\phi) C^{\sin\phi_S}_{UT}}
{\bar f_1(y_B)\,H_1(x_B)},\,
\\
&&D_{[\theta, \varphi]}=\frac{2 \sin^3\theta\cos 2\varphi}{1+\cos^2\theta},
\end{eqnarray}
where
$C^{\sin\phi_S}_{UT}=\Phi^{(1)}_{\gamma}(x_B)\, \widetilde\Phi^{(2)}_{\gamma}(\bar x_B)
\bar h(y_B)\hspace{-0.2cm} \stackrel{\hspace{0.2cm}\vec{\bf k}_{2}^{\,\perp}}{\circledast}
\hspace{-0.1cm}\varphi_{S}(y_B)$;
$\bar f_1(y_B)$ and $H_1(x_B)$ stem from the unpolarized cross-section and they parameterize the following matrix elements:
\begin{eqnarray}
\label{Unpol}
&&\langle P_2| \bar\psi\, \gamma^- \,\psi |P_2 \rangle\stackrel{{\cal F}}{\sim} P^-_2\bar f_1(y),\,
\nonumber\\
&&\langle P_1| \bar\psi_+ |q(K)\rangle \langle q(K)| \psi_+ |P_1 \rangle\stackrel{{\cal F}}{\sim} P_1^+ H_1(x),
\end{eqnarray}
where
\begin{eqnarray}
\label{Unpol2}
H_1(x)=
\frac{1}{2\bar x P_1^+} \int(d^2\vec{\bf k}_{1}^{\,\perp})
\varphi_{\gamma}^{(2)}(\bar x,\vec{\bf k}_{1}^{\,\perp})
\varphi_{\gamma}^{(1)}(x,\vec{\bf k}_{1}^{\,\perp}).
\nonumber
\end{eqnarray}
The new SSA defined by Eqn.~(\ref{NewSSA}) constitutes the main result of our study.

\section{Conclusion}

We derive the gauge invariant photon-induced DY hadron tensor with the essential
spin transversity and ``primordial'' transverse momenta.
In the paper, we focus on the case where one of photon distribution amplitudes has been projected onto the chiral-odd twist-$2$ combination
which singles out the chiral-odd parton distribution inside nucleons.

We predict a new single transverse spin asymmetry which can be probed experimentally
and which are associated with the spin transversity and the
nontrivial specific $\varphi$-angular dependence.
This asymmetry can be treated as a signal of
the gluon pole presence integrally with the study of both chiral-odd and chiral-even photon distribution amplitudes.

We emphasize that the possibility to study different SSAs induced by chiral-odd distribution functions/amplitudes appears only owing to the gluon pole presence.
Thus, the proposed angular dependence of SSA
can furnish the implicit evidences for the gluon pole observation in experiments.


{\bf Acknowledgments.} We thank A.V.~Efremov, L.~Motyka, O.V.~Teryaev and N.~Volchanskiy  for useful discussions.
The work by I.V.A. was supported by the Bogoliubov-Infeld Program.
I.V.A. also thanks the Theoretical Physics Division of NCBJ (Warsaw) for warm hospitality.
L.Sz. is supported by grant No No 2017/26/M/ST2/01074 of the National Science Center in
Poland.

\appendix
\renewcommand{\theequation}{\Alph{section}.\arabic{equation}}

\section{Demonstration of Separable Ansatz}
\label{App1}

In order to demonstrate the separable ansatz (\ref{ansatz}), it is instructive to
estimate the non-perturbative matrix element which corresponds to the distribution amplitude
within the frame of some effective model. A concrete realization of the model is not important
for our purposes. So, we begin with
the matrix element representing $\Phi^{(2)}_\gamma$ of the leading twist in the interaction representation.
We have
\begin{eqnarray}
\label{me1}
M&=&\int(d^4\eta) e^{i(\ell-k_2)\cdot\eta} \langle 0| T \bar\psi(0)\sigma^{+\perp}\psi(\eta) \mathbb{S} |P_1\rangle
\\
&=&\int(d^4\eta) e^{i(\ell-k_2)\cdot\eta} \int (d^4\xi) e^{-iP_1\cdot\xi}
\nonumber\\
&&\times
\langle 0| T \bar\psi(0)\sigma^{+\perp}\psi(\eta) \frac{\delta\mathbb{S}}{\delta M(\xi)} |0 \rangle,
\end{eqnarray}
where $\mathbb{S}$-matrix is given by the interaction action:
\begin{eqnarray}
\label{EffLag}
\mathbf{S}^{\text{eff.}}_I=g_M \int(d^4z) M(z) \bar\psi(z) \Gamma_M \psi(z).
\end{eqnarray}
Here, $M$ denotes the relevant meson which in our case is hadron-like photon.
Note that, within the effective frame, the coupling constant $g_M$ can be calculated with a help
of the bound (or coupling constant) condition (see for example \cite{Anikin:1995cf, Anikin:2000th}).
To the leading order of $g_M$, using the corresponding Fourier transformations, we obtain that
\begin{eqnarray}
\label{me2}
M=\text{tr}\left[ \sigma^{+\perp} S(\ell-k_2) \Gamma_M S(\ell-k_2-P_1) \right].
\end{eqnarray}
In the paper, we study the hadron-like behaviour of photon, therefore we can choose $\Gamma_M=\sigma^{+\perp}$.
\begin{widetext}
With this choice and within the frame (\ref{Sudakov-decom-1}), we have
\begin{eqnarray}
\label{me3}
M = \frac{\text{tr}\left[ \sigma^{+\perp} (\ell^+ \gamma^-) \sigma^{+\perp} ([\ell^+-P^+_1] \gamma^-) \right]}
{\left[(\ell-k_2)^2 + i0 \right]\left[(\ell-P_1-k_2)^2 + i0 \right]}
=\frac{\text{tr}\left[ \sigma^{+\perp} \gamma^- \sigma^{+\perp} \gamma^- \right]}
{\left[-2 k_2^- - (\vec{\ell}_\perp - \vec{\bf k}_2^\perp)^2/\ell^+ + i\epsilon \right]
\left[-2 k_2^- - (\vec{\ell}_\perp - \vec{\bf k}_2^\perp)^2/(\ell^+-P^+_1) + i\epsilon \right]}
\end{eqnarray}
Let us now focus on the typical function which can be extracted from Eqn.~(\ref{me3})
\begin{eqnarray}
\label{expr1}
\varphi((\vec{\ell}_\perp - \vec{\bf k}_2^\perp)^2)=\frac{1}
{\left[-2 k_2^- - (\vec{\ell}_\perp - \vec{\bf k}_2^\perp)^2/\ell^+ + i\epsilon \right]}.
\end{eqnarray}
Provided $|\vec{\ell}_\perp^{\,2}|\sim |\vec{\ell}_\perp\cdot\vec{\bf k}_2^\perp|
\sim |\vec{\bf k}_{2\,\perp}^{\,2}|\sim O(1)$, we can make an approximation
\begin{eqnarray}
\label{expr1}
&&\varphi((\vec{\ell}_\perp - \vec{\bf k}_2^\perp)^2)=
- \frac{1}{2k_2^-} \left[ 1 -
\frac{
(\vec{\ell}_\perp - \vec{\bf k}_2^\perp)^2}{2k_2^-\ell^+} + ... \right]
\approx - \frac{1}{2k_2^-} \left[ 1 -
\frac{
\vec{\ell}^{\,2}_\perp - \vec{\bf k}^2_{2\,\perp}}{2k_2^-\ell^+} + ... \right]
\approx - \frac{1}{2k_2^-} \left[ 1 -
\frac{\vec{\ell}^{\,2}_\perp}{2k_2^-\ell^+} + ... \right]
\left[ 1 +
\frac{\vec{\bf k}^2_{2\,\perp}}{2k_2^-\ell^+} + ... \right]
\nonumber\\
&&=- \frac{1}{2k_2^-} \varphi_1(\vec{\ell}^{\,2}_\perp) \varphi_2(\vec{\bf k}^2_{2\,\perp}).
\end{eqnarray}
This chain of approximations supports the separable ansatz.
\end{widetext}


\begin{thebibliography}{99}

\bibitem{Angeles-Martinez:2015sea}
  R.~Angeles-Martinez {\it et al.},
  Acta Phys.\ Polon.\ B {\bf 46}, 2501 (2015)

\bibitem{Boer:2011fh}
  D.~Boer {\it et al.},
  arXiv:1108.1713 [nucl-th].

\bibitem{Boer:2003cm}
  D.~Boer, P.~J.~Mulders and F.~Pijlman,
  Nucl.\ Phys.\ B {\bf 667}, 201 (2003)

\bibitem{Kang:2011hk}
  Z.~B.~Kang, J.~W.~Qiu, W.~Vogelsang and F.~Yuan,
  Phys.\ Rev.\ D {\bf 83}, 094001 (2011)

\bibitem{Boer:2011fx}
  D.~Boer,
  Phys.\ Lett.\ B {\bf 702}, 242 (2011)

\bibitem{Arnold:2008kf}
  S.~Arnold, A.~Metz and M.~Schlegel,
  Phys.\ Rev.\ D {\bf 79}, 034005 (2009)

\bibitem{Anikin:2010wz}
  I.~V.~Anikin and O.~V.~Teryaev,
  Phys.\ Lett.\ B {\bf 690}, 519 (2010)

\bibitem{Anikin:2015xka}
  I.~V.~Anikin and O.~V.~Teryaev,
  Eur.\ Phys.\ J.\ C {\bf 75}, no. 5, 184 (2015)

\bibitem{Anikin:2015esa}
  I.~V.~Anikin and O.~V.~Teryaev,
  Phys.\ Lett.\ B {\bf 751}, 495 (2015)

\bibitem{Anikin:2017azc}
  I.~V.~Anikin, L.~Szymanowski, O.~V.~Teryaev and N.~Volchanskiy,
  Phys.\ Rev.\ D {\bf 95}, no. 11, 111501 (2017)

\bibitem{Anikin:2017vvn}
  I.~V.~Anikin, L.~Szymanowski, O.~V.~Teryaev and N.~Volchanskiy,
  J.\ Phys.\ Conf.\ Ser.\  {\bf 938}, no. 1, 012065 (2017)

\bibitem{Anikin:2017pch}
  I.~V.~Anikin, L.~Szymanowski, O.~V.~Teryaev and N.~Volchanskiy,
  J.\ Phys.\ Conf.\ Ser.\  {\bf 938}, no. 1, 012039 (2017)


\bibitem{Bai:2013plv}
  M.~Bai {\it et al.}, ``Status and Plans for the Polarized Hadron Collider at RHIC,''

\bibitem{Baum:1996yv}
  G.~Baum {\it et al.} [COMPASS Collaboration],
  CERN-SPSLC-96-14, CERN-SPSLC-P-297.

\bibitem{COMPASS}
  M.~Aghasyan {\it et al.} [COMPASS Collaboration],
  arXiv:1704.00488 [hep-ex].

\bibitem{Kouznetsov:2017bip}
  O.~Kouznetsov and I.~Savin,
  Nucl.\ Part.\ Phys.\ Proc.\  {\bf 282-284}, 20.

\bibitem{Savin:2016arw}
  I.~Savin {\it et al.},
  Eur.\ Phys.\ J.\ A {\bf 52}, no. 8, 215 (2016).

\bibitem{Ioffe:1983ju}
  B.~L.~Ioffe and A.~V.~Smilga,
  Nucl.\ Phys.\ B {\bf 232}, 109 (1984).


\bibitem{Balitsky:1989ry}
  I.~I.~Balitsky, V.~M.~Braun and A.~V.~Kolesnichenko,
  Nucl.\ Phys.\ B {\bf 312}, 509 (1989).

\bibitem{Ball:2002ps}
  P.~Ball, V.~M.~Braun and N.~Kivel,
  Nucl.\ Phys.\ B {\bf 649}, 263 (2003)
  [hep-ph/0207307].

\bibitem{Braun:2002en}
  V.~M.~Braun, S.~Gottwald, D.~Y.~Ivanov, A.~Schafer and L.~Szymanowski,
  Phys.\ Rev.\ Lett.\  {\bf 89}, 172001 (2002)

\bibitem{Berger:1979du}
  E.~L.~Berger and S.~J.~Brodsky,
  Phys.\ Rev.\ Lett.\  {\bf 42}, 940 (1979).

\bibitem{Pire:2009ap}
  B.~Pire and L.~Szymanowski,
  Phys.\ Rev.\ Lett.\  {\bf 103}, 072002 (2009)

\bibitem{Barone:2001sp}
  V.~Barone, A.~Drago and P.~G.~Ratcliffe,
  Phys.\ Rept.\  {\bf 359}, 1 (2002)

\bibitem{Anikin:2016bor}
  I.~V.~Anikin, I.~O.~Cherednikov and O.~V.~Teryaev,
  Phys.\ Rev.\ D {\bf 95}, no. 3, 034032 (2017)

\bibitem{An-ImF}
  I.~V.~Anikin, D.~Y.~Ivanov, B.~Pire, L.~Szymanowski and S.~Wallon,
  Nucl.\ Phys.\  B {\bf 828}, 1 (2010)

\bibitem{Belitsky:2005qn}
  A.~V.~Belitsky and A.~V.~Radyushkin,
  Phys.\ Rept.\  {\bf 418}, 1 (2005)

\bibitem{Diehl:2003ny}
  M.~Diehl,
  Phys.\ Rept.\  {\bf 388}, 41 (2003)

\bibitem{Braun:2011dg}
  V.~M.~Braun and A.~N.~Manashov,
  JHEP {\bf 1201}, 085 (2012)

\bibitem{Anikin:1995cf}
  I.~V.~Anikin, M.~A.~Ivanov, N.~B.~Kulimanova and V.~E.~Lyubovitskij,
  Z.\ Phys.\ C {\bf 65}, 681 (1995).

\bibitem{Anikin:2000th}
  I.~V.~Anikin, A.~E.~Dorokhov, A.~E.~Maksimov, L.~Tomio and V.~Vento,
  Nucl.\ Phys.\ A {\bf 678}, 175 (2000).

\end{thebibliography}
\end{document}